\documentclass[prb,aps,twocolumn]{revtex4}

\usepackage{epsfig}
\usepackage{amsmath}

\begin{document}

\title{NMR study of the spin correlations in the $S=1$ armchair chain Ni$_2$NbBO$_6$}

\author{K. Y. Zeng$^{1}$}
\author{Long Ma$^{1}$}
\email{malong@hmfl.ac.cn}
\author{L. M. Xu$^{2}$}
\author{Z. M. Tian$^{2}$}
\email{tianzhaoming@hust.edu.cn}
\author{L. S. Ling$^{1}$}
\author{Li Pi$^{1,3}$}
\email{pili@ustc.edu.cn}

\affiliation{$^{1}$Anhui Province Key Laboratory of Condensed Matter Physics at Extreme
Conditions, High Magnetic Field Laboratory, Chinese Academy of Sciences, Hefei
230031, China\\
$^{2}$ School of Physics and Wuhan National High Magnetic Field Center, Huazhong University of Science and Technology, Wuhan 430074, PR China\\
$^{3}$ Hefei National Laboratory for Physical Sciences at the Microscale, University of Science and Technology of China, Hefei 230026, China}

\date{\today}

\begin{abstract}
We report our nuclear magnetic resonance (NMR) study on the structurally spin chain compound Ni$_2$NbBO$_6$ with complex magnetic coupling. The antiferromagnetic transition is monitored by the line splitting resulting from the staggered internal hyperfine field. The magnetic coupling configuration proposed by the first-principle density functional theory (DFT) is supported by our NMR spectral analysis. For the spin dynamics, a prominent peak at $T\sim35$ K well above the N\'{e}el temperature ($T_N\sim20$ K at $\mu_0H=10$ T) is observed from the spin-lattice relaxation data. As compared with the dc-susceptibility, this behavior indicates a antiferromagnetic coupling with the typical energy scale of $\sim3$ meV. Thus, the Ni$_2$NbBO$_6$ compound can be viewed as strongly ferromagnetically coupled armchair spin chains along the crystalline $b$-axis. These facts place strong constraints to the theoretical model for this compound.

\end{abstract}

\maketitle

The strong quantum fluctuations in the low-dimensional (low-D) antiferromagnets always leads to exotic quantum excitations as well as attractive quantum ground states\cite{Zapf_RMP, Han_nature, Punk_NP, Piazza_NP}. One of the archetype example is the spin-1/2 Heisenberg antiferromagnetic (AFM) chain system, which is proved exactly to host the quantum critical ground state called Tomonaga-Luttinger liquid\cite{Bethe_ZPhys_71_205}. For the spin excitation of such systems, the gapless "domain wall" excitation with fractional quantum number called spinon is theoretically proposed and further experimentally verified\cite{Faddeev_PLA_85_375}.

After the pioneering work of Haldane in 1983\cite{Haldane_PL_93A_464,Haldane_PRL_50_1153}, people begin to realize that chains with integer-spins (Haldane chains) are topologically different with that with half-odd spins. The ground state of Haldane chains is quantum disordered singlets, which can be viewed as a simple version of the valence bond solid state. In contrast with the gapless excitation continuum of the multi-spinon excitations, the excitations from singlets to triplets need overcome a finite energy gap (Haldane gap)\cite{Haldane_PL_93A_464,Haldane_PRL_50_1153}. Interestingly, the spin excitation spectrum of Haldane chains is asymmetric about the Brillouin zone boundary as a result of the maintaining translational symmetry of the crystalline lattice\cite{Ma_PRL_69_3571,Zaliznyak_PRL_87_017202,Bera_PRB_91_144414}, very different with the ordinary N\'{e}el ordered state. Up to now, several quasi-1D $S=1$ quantum magnets have been identified as ideal realization of the quantum singlet ground state\cite{Takigawa_PRB_52_R13087,Takigawa_PRL_76_2173,Pahari_PRB_73_012407,Mutka_PRL_67_497,Honda_PRB_63_064420,Uchiyama_PRL_83_632,Darriet_SSC_86_409}.

By further introducing the inter-chain coupling ($J_{\perp}$) and single-ion anisotropy ($D$) to the integer spin Hamiltonian, rich phase diagram can be reached in the $J_{\perp}-D$ space\cite{Wierschem_PRL_112_247203,Wierschem_PRB_86_201108}. Novel ground states, such as Bose-Einstein condensation of magnons\cite{Ruegg_PRL_93_257201,Wierschem_PRB_86_201108,Zapf_RMP}, spin supersolids\cite{Sengupta_PRL_99_217205,Sengupta_PRL_98_227201}, \textit{et al}. can be realized by tuning the competing interactions. By applied external magnetic field or physical pressure, quantum phase transitions can be triggered in such low-D spin systems. As a result, integer spin chain systems with complex structure and magnetic couplings has supplied valuable playgrounds for the condensed matter society to explore novel quantum excitations and criticality.

The AFM insulator Ni$_2$NbBO$_6$, first synthesized about four decades ago\cite{Ansell_ACSB_38_892}, exhibit complex lattice structure and magnetic interactions. The lattice structure can viewed as coupled armchair spin chains along the crystalline $b$-axis, or zig-zag spin chains along the $c$-axis, where the magnetic Ni$^{2+}$ sites carry integer $S=1$ spins. DC-susceptibility measurements indicate a AFM transition at $T_N\sim23.5$ K for a low magnetic field, and a field induced spin-flop transition near $\mu_0H\sim3.67$ T in the low temperature ordered phase, when the field is applied perpendicular to the armchair spin chain direction\cite{Rao_PRB_91_014423}. With density functional theory (DFT) calculations, the magnetic coupling configuration is determined, and a possible magnetic structure is proposed\cite{Rao_PRB_91_014423}. However, this result is contradict with the conventional Goodenough-Kanamori-Anderson (GKA) rules describing the magnetic interactions in insulators\cite{Goodenough_book,Streltsov_PU_60_1121,Prosnikov_PRB_98_104404}. From the Raman scattering study, most phonon peaks are identified, of which several modes exhibit strong spin-phonon coupling\cite{Prosnikov_PRB_98_104404}. For the magnetic scattering, three magnetic modes are observed, with the high-energy modes assigned to two-magnon modes\cite{Prosnikov_PRB_98_104404}.

Up to now, very little research into the spin correlations in Ni$_2$NbBO$_6$ is carried out, while at least two important questions still exist. One is about the complex coupling configuration, regarding the contradictory result shown above. The other one is related to whether Ni$_2$NbBO$_6$ should be treated as low-D antiferromagnets. Although this compound can be viewed as spin chain from the structural point of view, the temperature dependence of dc-susceptibility is very similar with ordinary 3D antiferromagnets. The broad peak in the susceptibility, typical for low-D antiferromagnets\cite{Law_JPCM_25_065601}, is completely absent here.

With nuclei as natural probe sitting on the lattice site, NMR is very useful in the study of the static magnetism as well as low-energy spin excitations in the quantum antiferromagnets. In this paper, we report study of the spin correlations in Ni$_2$NbBO$_6$ via $^{11}$B NMR spectroscopy as well as relaxation measurements. The AFM long-ranged order is indicated from the line splitting with magnetic field applied along $b$-axis. By the spectral analysis based on the lattice symmetry, this fact supports the magnetic coupling configuration previously proposed by the density functional theory (DFT) calculations instead of the GKA rules\cite{Rao_PRB_91_014423,Goodenough_book,Streltsov_PU_60_1121}. From the spin-lattice relaxation rates, a prominent peak at $T\sim35$ K is observed, which is well above the N\'{e}el temperature. In contrast with the temperature dependence of the Knight shift and $dc$-susceptibility, this peak originates from the short-ranged AFM correlations, which is the typical characteristic of quasi-1D AFM spin chains. Based on these observations, the Ni$_2$NbBO$_6$ compound can be viewed as strongly ferromagnetically coupled armchair $S=1$ spin chains along the crystalline $b$-axis.

\begin{figure}
\includegraphics[width=8.5cm, height=6.5cm]{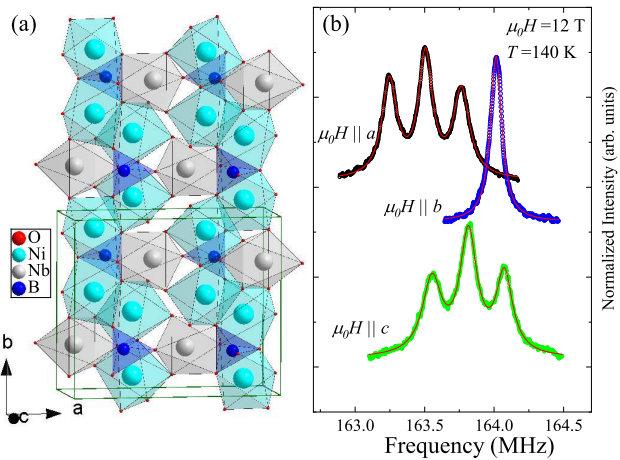}
\caption{\label{struc1}(color online) (a) The crystal structure of Ni$_2$NbBO$_6$ with distorted [NiO$_6$]/[NbO$_6$] octahedra and [BO$_4$] tetrahedra shown by the respective polyhedra. (b) Typical $^{11}$ NMR spectrum at the paramagnetic state ($T=140$ K) with a 12 Tesla field applied along three different crystalline axis. The red lines are fittings to the data by the Lorentz peak function.
}
\end{figure}

Single crystals of Ni$_2$NbBO$_6$ are synthesized with the conventional flux method as described elsewhere\cite{Ansell_ACSB_38_892,Rao_PRB_91_014423}. Dark green crystals with typical dimensions of $1.5\times1.5\times1$ mm$^3$ are chosen for our NMR measurements. The crystal directions are identified by single-crystal x-ray diffraction. The precise alignment of the magnetic field with the crystalline axis is guaranteed by placing the sample on a piezoelectric nanorotation stage. Our NMR measurements are conducted on the $^{11}$B nuclei($\gamma_n/2\pi=13.655$ MHz/T, $I=3/2$) with a phase coherent NMR spectrometer.
The $^{11}$B NMR spectra are obtained by summing up the frequency-swept spin-echo intensities. The spin-lattice relaxation rates are measured by the conventional inversion-recovery pulse sequence, and fitting the nuclear magnetization to the standard recovery function for nuclei with $I=3/2$.

\begin{figure}
\includegraphics[width=8.5cm, height=6.5cm]{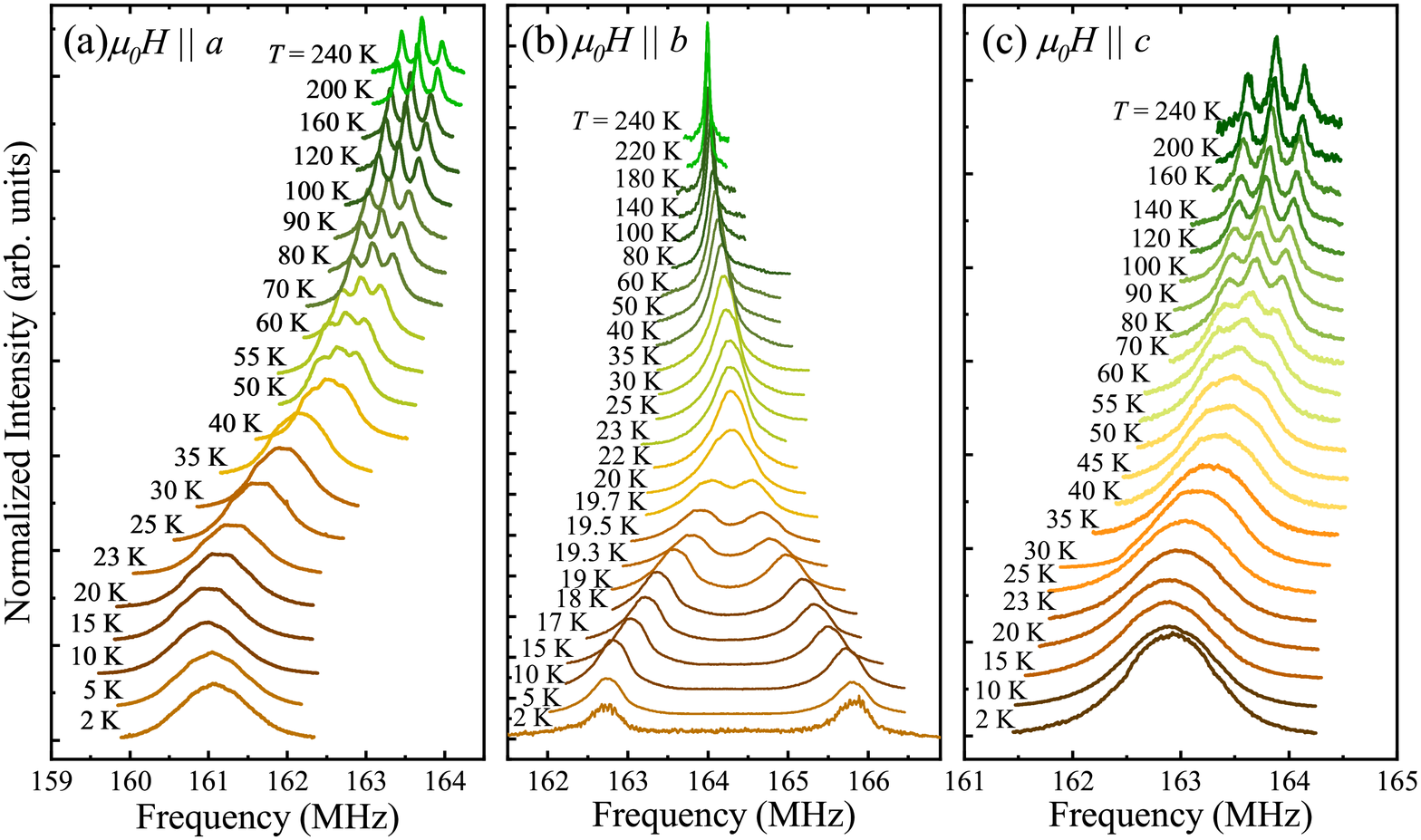}
\caption{\label{spec2}(color online)
The $^{11}$B NMR spectra at different temperatures with a 12 tesla field applied along the crystalline $a$-axis (a), $b$-axis (b) and $c$-axis (c).
}
\end{figure}

The Ni$_2$NbBO$_6$ crystalizes in the orthorhombic structure with the $pnma$ space group\cite{Ansell_ACSB_38_892}. The lattice structure is sketched in Fig.\ref{struc1}(a). Along the crystalline $b$-axis, double edge-shared [NiO$_6$] octahedra are linked by the [NbO$_6$] octahedra and [BO$_4$] tetrahedra, forming the armchair shaped spin chain. Alternatively, the lattice can also be viewed as zigzag spin chain along the $c$-axis. Only one position of $^{11}$B-site is present in the lattice. The [BO$_4$] tetrahedra share their corners with the [NiO$_6$] octahedra, leading to the strong coupling between the $^{11}$B nuclei and the spins located on Ni$^{2+}$, and making the $^{11}$B-NMR a very sensitive local probe of the magnetism.

We start to discuss our NMR results by presenting typical frequency-swept NMR spectra with the magnetic field applied along three different crystalline axis in Fig.\ref{struc1}(b). For fields along $a$- or $c$-axis, three resonance peaks are identified, with one central peak and double satellite peaks symmetrically located at both sides. Applying the field along the $b$-axis, we observe only one sharp peak. All the peaks can be well reproduced by the lorentz peak function as shown by the fitting lines.

The Hamiltonian for the quadrupole nuclei system with the nuclear spin $\bf{I}$ and quadrupole moment $Q$ in the presence of the applied magnetic field $\mathbf{H}$ can be generally written as\cite{Slichter_book,Abragam_book,Vachon_JPCM_20_295225},
\[
\begin{split}
H=-\gamma_n\hbar(1+K)\mathbf{H}\cdot\mathbf{I}+\frac{e^2qQ}{4I(2I-1)}(3I_Z^2-I(I+1)\\
+\frac{1}{2}\eta(I_+^2+I_-^2)).
\end{split}
\]
Here $K$ is the Knight shift defined as the relative line shift with respective to the nuclear Larmor frequency. The $X$, $Y$, $Z$ directions are the principle axis of the electric field gradient (EFG) tensor \{${V_{ij}}$\}, and the $\eta$ formulated by $\eta\equiv(V_{XX}-V_{YY})/V_{ZZ}$ is the in-plane anisotropy of the EFG. In the high magnetic field regime, the resonance frequency can be calculated by the first order pertubation theory,
\[
\begin{split}
\omega=-\gamma_n(1+K)H_0+\omega_Q(m-1/2)\\
\times(3\cos^2\theta-1+\eta\sin^2\theta\cos2\phi),
\end{split}
\]
where the quadrupolar frequency $\omega_Q$ equals to $3e^2qQ/(\hbar2I(2I-1))$, and $m$ is the magnetic quantum number.
As a result, one can expect to observe three transitions for $^{11}$B nuclei with $I=3/2$ for most magnetic field angles, while only one transition for some special angles (here is $\mu_0H||b$-axis).

In Fig.\ref{spec2}, we show $^{11}$B NMR spectra at different temperatures with the field applied along three different crystalline axis. With the magnetic field applied parallel with the $a$- or $c$-axis (Fig.\ref{spec2} (a) and (c)), the temperature dependence of $^{11}$B spectrum shares a similar behavior. All the three peaks shift to the low frequency side with the sample cooling down for the paramagnetic state ($T>T_N=20$ K). When the spin system enters the AFM state, the spectral frequency tends to become temperature independent. For the $\mu_0H||b$-axis case(Fig.\ref{spec2} (b)), the single peak first shift to the high frequency side above $T_N$, and splits symmetrically into double peaks in the AFM ordered state. For all the three field orientations, the spectra broadens gradually upon cooling.

\begin{figure}
\includegraphics[width=8.5cm, height=7.5cm]{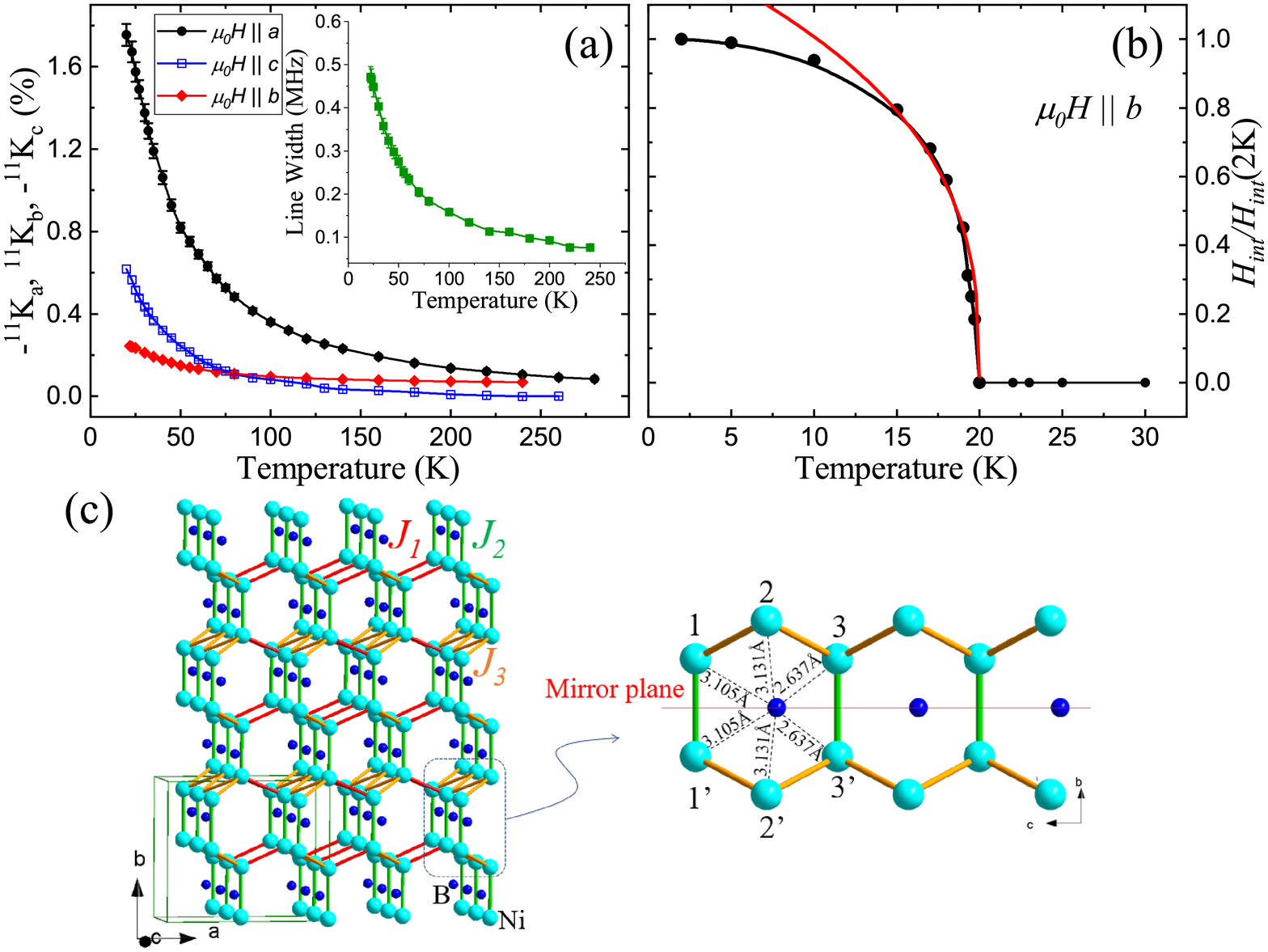}
\caption{\label{afm3}(color online)
  (a) The temperature dependence of the Knight shift for different magnetic field orientations. The Knight shift data $^{11}K$ with $\mu_0H||a$ and $\mu_0H||c$ is multiplied by $-1$ to get rid of the influence of the minus hyperfine coupling constant. Inset: The line width as a function of temperature with $\mu_0H||b$. (b) The internal field calculating from the frequency gap of the splitting peaks as compared with that at $T=2$ K versus temperature. (c) The simplified crystalline structure with the magnetic coupling $J_1$, $J_2$ and $J_3$ marked. Only the $^{11}$B sites and the magnetic Ni$^{2+}$ sites are shown for clarity. The six adjacent Ni$^{2+}$ magnetic sites with respect to the $^{11}$B sites are shown in the righthand side, with the distance between them marked respectively.
}
\end{figure}

To study the spin susceptibility and the magnetic interactions, we plot the temperature dependence of the Knight shift and the internal field for the AFM state respectively in Fig.\ref{afm3} (a) and (b). The Knight shift is a good measure of the spin susceptibility with the general expression of\cite{Vachon_JPCM_20_295225}
$$\mathbf{K}=\{A_{ij}\}\mathbf{\chi_s}(\mathbf{q}=0).$$
The tensor \{$A_{ij}$\} is the coupling constant between the nuclear and electronic spins. The second order correction to the resonance frequency of the central transition is very small, thus is neglected in the calculation of Knight shifts. For the present sample with $\mu_0H||a-$ or $c-$ axis, the element of the coupling tensor happens to be negative, resulting the Knight shifts with negative values. The temperature dependence of $|K|$ (Fig.\ref{afm3} (a)) show a typical upturn behavior upon cooling, indicating the enhancement of the spin susceptibility. This is consistent with the $dc$-susceptibility measurements. With the field along $b$-axis, we can precisely determine the line width of the only resonance peak (shown in Fig.\ref{afm3} (a) inset). The temperature dependence of the line width also shows an upturn behavior at low temperatures, which can be well scaled with the Knight shift. Thus, the line broadening originates from the enhanced Knight shift at low temperatures.

Next, we study the magnetic interactions via the spectral analysis below $T_N$. Line splitting due to the setup of magnetic ordering is observed below $T_N$ with the magnetic field applied along $b$-axis. In the AFM state, the magnetic unit cell is doubled as compared with the crystalline one.
One can expect a non-zero staggered internal field on the nuclear sites. When the external field is applied along the direction of the internal field, the total field can be calculated as $\mathbf{H}_{tot}=\mathbf{H}_{ext}+\mathbf{H}_{int}$, leading to the line splitting in the AFM state\cite{Casola_PRB_86_165111,Nath_PRB_80_214430}. In the present sample, the line splitting with $\mu_0H||b$-axis indicates the staggered internal field along this crystalline direction. The measured internal field serving as the order parameter of the AFM transition is plotted against temperature in Fig.\ref{afm3} (b). By fitting the data near $T_N$ to the function $H_{int}\propto (1-T/T_N)^{\beta}$, the critical exponent $\beta$ is determined to be $\sim0.35$. This is consistent with the 3D characteristic of the AFM transition\cite{Nath_PRB_80_214430,Pelissetto_PR_368_549}.

Our data support the magnetic coupling configuration proposed by the DFT calculations\cite{Rao_PRB_91_014423}, instead of the GKA rules. We show the simplified crystal structure in Fig.\ref{afm3}, only presenting the magnetic Ni$^{2+}$ and $^{11}$B sites. The nearest-neighbour, next-nearest-neighbour and next-next-nearest-neighbour magnetic coupling is marked with $J_1$, $J_2$ and $J_3$, respectively. For every $^{11}$B, six adjacent Ni$^{2+}$ magnetic moments labelling as $1$, $1^{\prime}$, \textit{et al}. contribute to the internal field (See the enlarged version). Locally, the mirror symmetry about the plane perpendicular to the $b$-axis is maintained. In the magnetically ordered state, the internal field contributed by $1$-sites can be generally written as\cite{Kitagawa_JPSJ_77_114709}
$$
\left(
  \begin{array}{c}
          H_{int}^a \\
          H_{int}^b \\
          H_{int}^c
 \end{array}
\right)
=
\left(
  \begin{array}{ccc}
          A_{aa}\quad A_{ab}\quad A_{ac}\\
          A_{ba}\quad A_{bb}\quad  A_{bc}\\
          A_{ca}\quad A_{cb}\quad  A_{cc}
 \end{array}
\right)
\left(
  \begin{array}{c}
          M^a \\
          M^b \\
          M^c
 \end{array}
\right),
$$
where the 2-ranked tensor $\{A_{ij}\}$ describes the coupling between $^{11}$B nuclear spins and the magnetic moments located on Ni$^{2+}$ $1$-sites.
Under the mirror symmetry operation, the coupling tensor between $^{11}$B and the $1^{\prime}$-sites is obtained to be
$$
\left(
  \begin{array}{ccc}
          A_{aa}\quad -A_{ab}\quad A_{ac}\\
          -A_{ba}\quad A_{bb}\quad  -A_{bc}\\
          A_{ca}\quad -A_{cb}\quad  A_{cc}
 \end{array}
\right).
$$
As proposed by DFT calculations\cite{Rao_PRB_91_014423}, the $J_2$ coupling is AFM, while $J_1$ and $J_3$ couplings are ferromagnetic. Because the magnetic frustration can be neglected in this compound\cite{Rao_PRB_91_014423}, the magnetic structure at low temperatures is determined by the magnetic interactions. Based on this configuration, the internal field at the $^{11}$B sites contributed from the $1$- and $1^{\prime}$-sites can be calculated as,
$$
\left(
  \begin{array}{c}
          H_{int}^a\\
          H_{int}^b\\
          H_{int}^c
 \end{array}
\right)
=
\left(
  \begin{array}{c}
          2A_{ab}M^b\\
          2A_{ba}M^a+2A_{bc}M^c\\
          2A_{cb}M^b
 \end{array}
\right).
$$
For both $2$- and $2^{\prime}$-sites and $3$- and $3^{\prime}$-sites, similar results can be obtained.

From the temperature dependence of $dc$-suspcetibility and the field induced spin-flop transition with the field along $c$-axis, the ordered moment aligned with $c$-axis can be easily determined in our sample, which is different from the reported $a$-axis\cite{Rao_PRB_91_014423}. However, this inconsistency does not affect the analysis here. Thus, one can imagine the staggered internal fields along the $b$-axis for $^{11}$B nuclei stacking on different positions along $b$-axis. As the $M^b$ component is zero, the line-splitting with $\mu_0H||a$-axis is absent. This is fully consistent with our NMR results. For the strong NMR magnetic field along $c$-axis, the line-splitting is also absent, further suggest that the magnetic moments flop to the crystalline $a$-axis in our sample.

Our observations contradict with the magnetic coupling configuration indicated by GKA rules\cite{Goodenough_book,Streltsov_PU_60_1121,Prosnikov_PRB_98_104404}. According to the GKA rules for the superexchange magnetic coupling in insulators, the sign of interaction strongly depend on the bond angle between the magnetic sites and the intermediate ligands. From the lattice structure, the $J_3$-coupling is determined to be AFM, while the $J_1$ and $J_2$-couplings are FM. Based on this configuration, the internal field contributed from $1$- and $1^{\prime}$-sites is
$$
\left(
  \begin{array}{c}
          H_{int}^a\\
          H_{int}^b\\
          H_{int}^c
 \end{array}
\right)
=
\left(
  \begin{array}{c}
          2A_{aa}M^a+2A_{ac}M^c\\
          2A_{bb}M^b\\
          2A_{ca}M^a+2A_{cc}M^c
 \end{array}
\right).
$$
This is contradict with the line-splitting with $\mu_0H||b$-axis observed in our NMR data. The failure of GKA rules in the present sample may be  related with the coupling of other side groups with the intermediate ligands\cite{Geertsma_PRB_54_3011}, which also imply the complicated magnetic coupling in the sample.

\begin{figure}
\includegraphics[width=8.5cm, height=7.5cm]{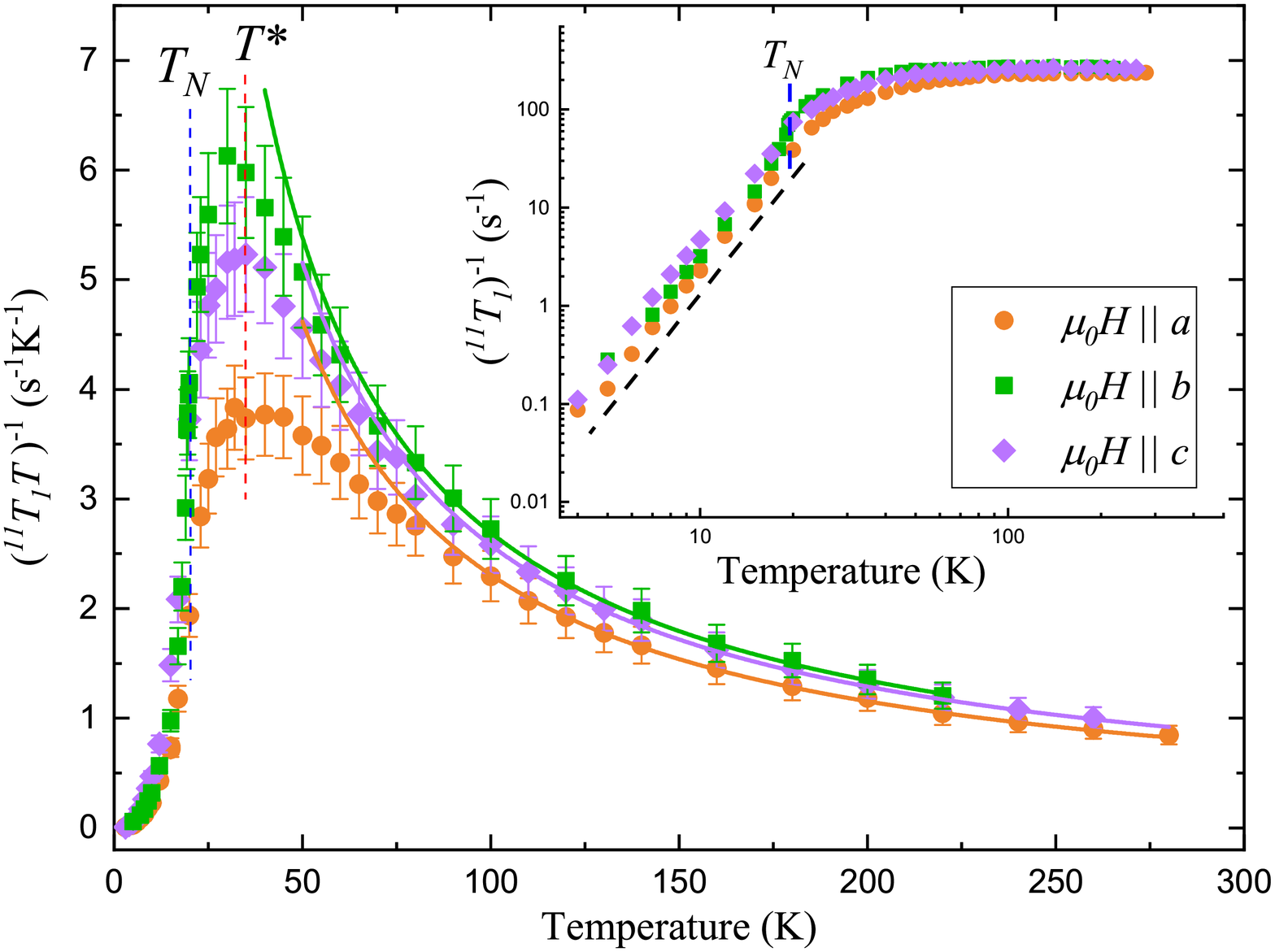}
\caption{\label{slrr4}(color online)
  The spin-lattice relaxation rate divided by temperature $(^{11}T_1T)^{-1}$ plotted versus temperature for different field directions. The solid lines are fittings to the Curie-Weiss law (See the main text). Inset: The temperature dependence of $T_1T$.
}
\end{figure}

Next, we focus on the question whether Ni$_2$NbBO$_6$ can be viewed as low-D magnets. For the AFM chain system, a typical broad peak in the temperature dependence of the susceptibility can be expected, which results from the build-up of short-range correlations at low temperatures\cite{Law_JPCM_25_065601}. However, the susceptibility of Ni$_2$NbBO$_6$ follows a well-defined Curie-Weiss upturn behavior from room temperature down to $T_N$, similar with other ordinary 3D antiferromagnets. We study the spin excitations through the spin-lattice relaxation measurements. The spin-lattice relaxation rate (SLRR) formulated as\cite{Slichter_book,Abragam_book},
$$
(T_1)^{-1}\propto T\sum_{\overrightarrow{q}}|A(\overrightarrow{q})|^2\frac{\chi^{''}(\overrightarrow{q},\omega_L)}{\omega_L},
$$
is a good probe of the spin-correlations in solids, where the $\chi^{''}(\overrightarrow{q},\omega_L)$ is the imaginary part of the dynamic spin susceptibility at the Larmor frequency $\omega_L$. By dividing the temperature, $(T_1T)^{-1}$, is the sum of the spin correlations in the reciprocal space.

In Fig.\ref{slrr4}, we present the temperature dependence of $(T_1T)^{-1}$ for three different field directions. For $\mu_0H||a$- and $c$-axis, the $(T_1)^{-1}$ is obtained by fitting the nuclear magnetization recovery curve to the standard function for the $1/2\leftrightarrow-1/2$ transition of the nuclei with $I=3/2$\cite{Mcdowell_JMR_113_242},
$$
\frac{M(t)}{M(\infty)}=1-f[0.9\exp(\frac{-t}{T_1})+0.1\exp(\frac{-t}{6T_1})].
$$
For $\mu_0H||b$-axis, the single exponential magnetization recovery function is used as all the three transitions are excited in the experiment. In the present single crystal, all the fitting curves are perfect without any stretching or other different $T_1$-components. This further indicate the uniform excitation behavior in our sample. With the sample cooling from high temperature, the $(T_1T)^{-1}$ shows an upturn behavior which can be described by the Curie-Weiss law, $(T_1T)^{-1}\propto1/(T+\theta)$ (shown by the fitting lines in Fig.\ref{slrr4}). When the temperature is further lowered, the $(T_1T)^{-1}$ begin to deviate from the Curie-Weiss law, and start to drop at $T^*\sim35$ K, well above the AFM transition $T_N=20$ K. With the sample entering the AFM long-range-ordered state, the $(T_1T)^{-1}$ show a typical power-law temperature dependence, which is clearly shown by the dashed in Fig.\ref{slrr4} inset.

The spin excitation nature seen from $(T_1T)^{-1}(T)$ support the low-D characteristic of the magnetism in Ni$_2$NbBO$_6$. For the high temperature region, the Curie-Weiss upturn behavior in $(T_1T)^{-1}$ is fully consistent with the the $dc$-susceptibility mainly measuring the spin correlations at $\overrightarrow{q}=0$, suggesting the weak $\overrightarrow{q}$-dependence of the spin excitation. In the low temperature region below $T_N$,
the power-law temperature dependence of $(T_1T)^{-1}$ is contributed from the spin-wave excitations of the 3D ordered state\cite{Beeman_PR_166_359}. The $(T_1T)^{-1}$ measures the sum of the spin correlations in the reciprocal space, while the $dc$-susceptibility and Knight shift is contributed by the spin excitations at $\overrightarrow{q}=0$. As a result, the contrasting prominent peak of $(T_1T)^{-1}$ at $T^*=35$ K indicates the AFM short-ranged spin correlations with the energy scale of $\sim 3$ meV, which is the fingerprint of the spin excitations in low-D spin chain systems\cite{Law_JPCM_25_065601}. The observed energy scale of $J_2$ is very close to the estimation of $J_2\sim 2.43$ meV by DFT calculations\cite{Rao_PRB_91_014423}. As the 3D ordering temperature is comparable to this energy scale, the spin chain system in Ni$_2$NbBO$_6$ should lies on the 1D-3D crossover regime.

To conclude, we have carried out detailed NMR study into the spin correlations in Ni$_2$NbBO$_6$ compound. The AFM long-ranged order is monitored by the line splitting with a magnetic field along $b$-axis. By the spectral analysis based on local crystalline symmetry, supported is the magnetic coupling configuration proposed by DFT calculations. The deviation from the general GKA rules is proposed to originate from the coupling of other side groups with the intermediate ligands. From the spin-lattice relaxation rates, a prominent broad peak at $T^*=35$ K is observed in the temperature dependence of $(T_1T)^{-1}$. This behavior is contributed from the short-ranged AFM correlations with a typical energy scale of $\sim3$ meV, further indicating the low-D characteristic of the magnetic behavior in Ni$_2$NbBO$_6$. As a result, Ni$_2$NbBO$_6$ can be viewed as a strongly coupled armchair spin chain system along $b$-axis lying on the 1D-3D crossover regime, which have placed strong constraints on the theoretical models describing this material.

This research was supported by the National Key Research and Development Program of China (Grant No. 2016YFA0401802), the National Natural Science Foundation of China (Grants No. 11874057, 11504377, 11574288, 11874158, U1732273 and 21927814) and the Users with Excellence Program of Hefei Science center CAS (Grant No. 2019HSC-UE008). A portion of this work was supported by the High Magnetic Field Laboratory of Anhui Province.

%\bibliography{NNBO}

\end{document}